\documentclass[useAMS,usenatbib]{mn2e}
\usepackage{graphicx}

\usepackage[update,prepend]{epstopdf}

\title[Polar caps]{Polar caps in the presence of an induction field}
\author[P. B. Jones]{P. B. Jones\thanks{E-mail:
p.jones1@physics.ox.ac.uk}  \\
University of Oxford, Department of Physics, Denys Wilkinson Building,\\
Keble Road, Oxford OX1 3RH, U.K.}

\begin{document}

\date{}

\pagerange{\pageref{firstpage}--\pageref{lastpage}} \pubyear{}

\maketitle

\label{firstpage}

\begin{abstract}

Following the early paper of Goldreich \& Julian (1969), polar-cap models have usually assumed that the closed sector of a pulsar magnetosphere corotates with the neutron star.  Recent work by Melrose \& Yuen has been a reminder that in an oblique rotator, the induction field arising  from the time-varying magnetic flux density cannot be completely screened.  The principal consequence is that the plasma does not corotate with the star.  Here it is shown that the physics of the polar cap is not changed at the altitudes of the radio emission source.  But the presence of a plasma drift velocity in the corotating frame of reference does provide a mechanism whereby the net charge of the star can be maintained within a stable band of values. It also shows directly how electron injection and acceleration occur in the outer gap of the magnetosphere.  It is consistent with radio-loud pulsars in the Fermi LAT catalogue of $\gamma$-emitters all having positive polar-cap charge density.

\end{abstract}


\section{Introduction}

The present consensus is that the source of the universal coherent radio emission in pulsars is in the open sector of the magnetosphere at low altitudes above the polar cap. In this region, the ${\bf B}$-field is so large that it must be regarded as fixed and model-independent, and basic physical considerations can be used to determine the nature of the acceleration field ${\bf E}_{\parallel}$ and the composition of the plasma. (Throughout this paper, the parallel and perpendicular directions are with respect to the local magnetic flux density ${\bf B}$.)  The assumption of plasma corotation in the closed sector of the magnetosphere has been an essential feature of polar-cap models, and follows from the paper of Goldreich \& Julian (1969) which treated the aligned rotating neutron star with no induction component, ${\bf E}^{ind} = 0$, in the observer-frame electric field.  In this case, the electric field generated by the rotation is exactly screened by the Goldreich-Julian charge density $\rho_{GJ}$.

The presence of an induction field in an observable pulsar whose magnetic moment is at an angle $\psi$ with the rotation angular velocity ${\bf \Omega}$ has usually not been the subject of comment.  But two recent papers by Melrose \& Yuen (2012, 2014) have examined the consequences of the induction field given directly in the observer frame, in terms of the time-dependence of ${\bf B}$, by Faraday's law.  They express the field as ${\bf E}^{ind} = {\bf E}^{ind}_{\parallel} + {\bf E}^{ind}_{\perp}$ and consider the screening of these components separately. The parallel component can, in principle, be screened precisely by a charge density $\tilde{\rho}$ whilst the perpendicular component causes a plasma drift velocity in the corotating frame. They note that non-corotation must affect the distinction between open and closed sectors of the magnetosphere.  The consequences of this for polar-cap physics and for the outer gap of the magnetosphere are the subject of the present brief paper.

A further problem that has been the subject of little comment is the question of how currents flowing from the polar cap return so as to maintain the net charge of the star within a stable and limited band of values.  Polar-cap models have ignored this problem, though its treatment is implicit in force-free models of the outer magnetosphere. In the computational aspects of this latter work, the neutron-star radius was artificially large, approximately $0.2R_{LC}$, where $R_{LC}$ is the light cylinder radius, and the boundary condition assumed on the neutron-star surface is an equipotential with space-charge-limited flow.  The force-free solutions specify the charge and current densities at all points in the magnetosphere.  We refer to Bai \& Spitkovsky (2010) for some informative computed distributions of these quantities.  

The assumption of corotation in the closed magnetosphere creates problems in understanding, for example in a polar-cap ${\bf \Omega}\cdot{\bf B} < 0$ pulsar, how the outflow of positive charge in the open sector can be compensated by a flux of electrons.  However, the plasma drift calculated by Melrose \& Yuen does obviate this difficulty and enables us to see, at least qualitatively, how the necessary charge stability can be maintained.

We refer to Melrose \& Yuen (2012, 2014) for a full discussion of the induction field and related topics but give here in Section 2 a very brief summary of the relevant results.  These are considered in relation to the closed magnetosphere in Section 3, the polar cap and coherent emission processes in Section 4, and the outer gap in section 5.

Unless otherwise stated, neutron stars with polar-cap ${\bf \Omega}\cdot {\bf B} < 0$ are assumed.  The reason for this is that the polar-cap flux of positive particles has been shown to be a non-steady-state plasma, principally of ions and protons. The acceleration field ${\bf E}_{\parallel}$ is limited
by the reverse flux of electrons produced by ion motion through blackbody radiation. The accelerated plasma is relativistic but not ultra-relativistic and is an ideal basis for the growth of Langmuir modes and turbulence. Pair creation is not significant except at very high magnetic flux densities in which case it actually suppresses growth of the Langmuir mode.  The properties of the mode do not otherwise depend on the polar magnetic field and so differ little over a wide range of objects, from normal pulsars to millisecond pulsars (MSP) and are consistent with the almost universal spectral properties of the radio emission.  We refer to Jones (2015) and papers cited therein for further details of the case for ${\bf \Omega}\cdot{\bf B} <0$ pulsars. Possible observable consequences of plasma drift in the corotating frame for the ${\bf \Omega}\cdot{\bf B} > 0$ case are considered briefly in Section 6.

\section{The induction field and screening}

The induction field ${\bf E}^{ind}$ is given directly in the observer frame, in terms of the time-dependence of ${\bf B}$, by Faraday's law.  It is divergence-free, $\nabla\cdot{\bf E}^{ind} = 0 $, and so cannot be universally screened by the magnetosphere plasma. The parallel component can be screened completely, in principle, by a time-dependent charge density
$\tilde{\rho} = - \nabla\cdot{\bf E}^{ind}_{\parallel}/(4\pi)$ provided the plasma contains an adequate density of charged particles. For the dipole field assumed, this can be expressed as,
\begin{eqnarray}
\tilde{\rho} = -\rho_{GJ0}\left(\cos\theta - \cos\theta_{m}\cos\psi\right)\frac{9\cos\theta_{m}(1 + \cos^{2}\theta_{m})}{2(1 + 3\cos^{2}\theta_{m})^{2}},
\end{eqnarray}
(Melrose \& Yuen 2014) in which $\theta$ and $\phi$ are observer-frame spherical polar coordinates defined with respect to the rotation axis and $\rho_{GJ0}(\eta) = \Omega B/2\pi c$ is the Goldreich-Julian charge density defined for $\theta = 0$ and $\psi = \pi$, where ${\bf \Omega}\cdot{\bf B} =\Omega B\cos\psi$.  Here $\eta$ is the polar coordinate in units of the neutron star radius $R$. Equation (1) is stationary in the corotating frame.  The magnetic colatitude $\theta_{m}$ is,
\begin{eqnarray}
\cos\theta_{m} = \cos\psi\cos\theta \pm \sin\psi\sin\theta\cos(\phi - \omega t),
\end{eqnarray}
in which the appropriate sign is that of the polar-cap ${\bf \Omega}\cdot{\bf B}$.
The general value of the Goldreich-Julian density is,
\begin{eqnarray}
\rho_{GJ} = -\frac{1}{2} \rho_{GJ0}\left(3\cos\theta\cos\theta_{m} - \cos\psi\right).
\end{eqnarray}
The component ${\bf E}^{ind}_{\parallel}$ is then completely removed.  (The consequent polarization current density given by the continuity equation exactly replaces the displacement current component $\partial{\bf E}^{ind}_{\parallel}/c\partial t$ in Maxwell's equation.)  We shall adopt the approximation that $\tilde{\rho}$ can be treated as a perturbation to the aligned corotating state (for small $\psi$, see Melrose \& Yuen 2014) so that the total screening charge density is $\rho_{s} = \rho_{GJ} + \tilde{\rho}$.

Melrose \& Yuen show that the effect of the component ${\bf E}^{ind}_{\perp}$ on the plasma is quite different.  In the plane locally perpendicular to ${\bf B}$, the plasma has a drift velocity $\tilde{\bf v} = c{\bf E}^{ind}_{\perp}\times{\bf B}/B^{2}$ relative to the corotating frame. The polarization current density in this plane is perpendicular to the drift velocity. Owing to the very large values of $B$ for a neutron-star plasma, the Alfv\'{e}n velocity is $v_{A} \gg c$ so that this component of the polarization current density is negligible and can give no significant screening of ${\bf E}^{ind}_{\perp}$. The drift velocity given by Melrose \& Yuen (2012) for dipole field geometry is precisely azimuthal on the magnetic axis so that in the vicinity of the polar cap we require only this component.  Directly from these authors' equation (10) it is,
\begin{eqnarray}
\tilde{v}_{\phi} = \pm \frac{2\Omega R\eta\sin\psi\cos\theta_{m}\cos(\phi - \Omega t)}
{1 + 3\cos^{2}\theta_{m}},
\end{eqnarray}
relative to the corotating frame (Melrose \& Yuen 2012) in which it is time-independent. Again the correct sign is that of ${\bf \Omega}\cdot{\bf B}$. The degree to which polar-cap physics is affected by $\tilde{v}_{\phi}$ is described in Sections 3 and 4.

\section{The closed magnetosphere}

The closed magnetosphere has an atmosphere close to local thermodynamic equilibrium.  Above that atmosphere, it is divided into sectors of positive and negative screening charge density separated by a null surface. The neutrality of its atmosphere requires that the electron and proton chemical potentials satisfy the relation $\mu_{e}(z) - \mu_{e}(0) = \mu_{p}(z) - \mu_{p}(0)$ at altitudes $z$ within it, and this is achieved through the presence of a small non-zero $E_{\parallel}$ which balances the surface gravitational acceleration $g_{s}$,
\begin{eqnarray}
{\rm e}E_{\parallel} = - (m_{p} - m_{e})\frac{g_{s}}{2}, \hspace{1cm} g_{s} < 0.
\end{eqnarray}
The scale height is of the order of $10^{-1}$ cm and the field within it is $E_{\parallel} \approx 100$ V cm$^{-1}$.  The boundary condition at $z = 0$ allows free transit of electrons, protons or positive ions so that electric fields which do not allow an exponentially decreasing number density of either electrons or protons at the top of the atmosphere are given by ${\rm e}E_{\parallel}(z) < m_{e}g_{s}$ or ${\rm e}E_{\parallel}(z) > - m_{p}g_{s}$, respectively.

There are two reasons why such an electric field component should exist from time to time to disturb the local thermodynamic equilibrium.  Firstly, the screening charge density $\rho_{s}$ specified by equations (1) and (3) is stationary in the corotating frame but is a function of $\phi$ and has to be supplied by plasma drifting locally perpendicular to ${\bf B}$ with velocity $\tilde{\bf v}$.  This azimuthal variation of $\rho_{s}$ is shown in Figure 1: the extent to which it is significant is an increasing function of $\psi$. This takes no account of the additional charge density which is required to maintain the open-closed sector boundary condition. The correct charge density on a flux line at any instant has to be maintained by transit of electrons or protons from the surface atmosphere and a combination of drift and motion  parallel with ${\bf B}$.  Also, the flux of protons and ions (in an ${\bf \Omega}\cdot{\bf B} < 0$ pulsar) in the open sector leads to a rate change of the order of $\partial E_{\parallel}/\partial t\sim -\pi u_{0}^{2}\rho_{GJ0}c/2R^{2}$, in which $u_{0}$ is the open polar-cap radius.  This rate of change in $E_{\parallel}$  must also be compensated by electron transit at $z = 0$. It would be present even if there were no induction field.

\begin{figure}
\includegraphics[trim=20mm 30mm 40mm 130mm, clip, width=84mm]{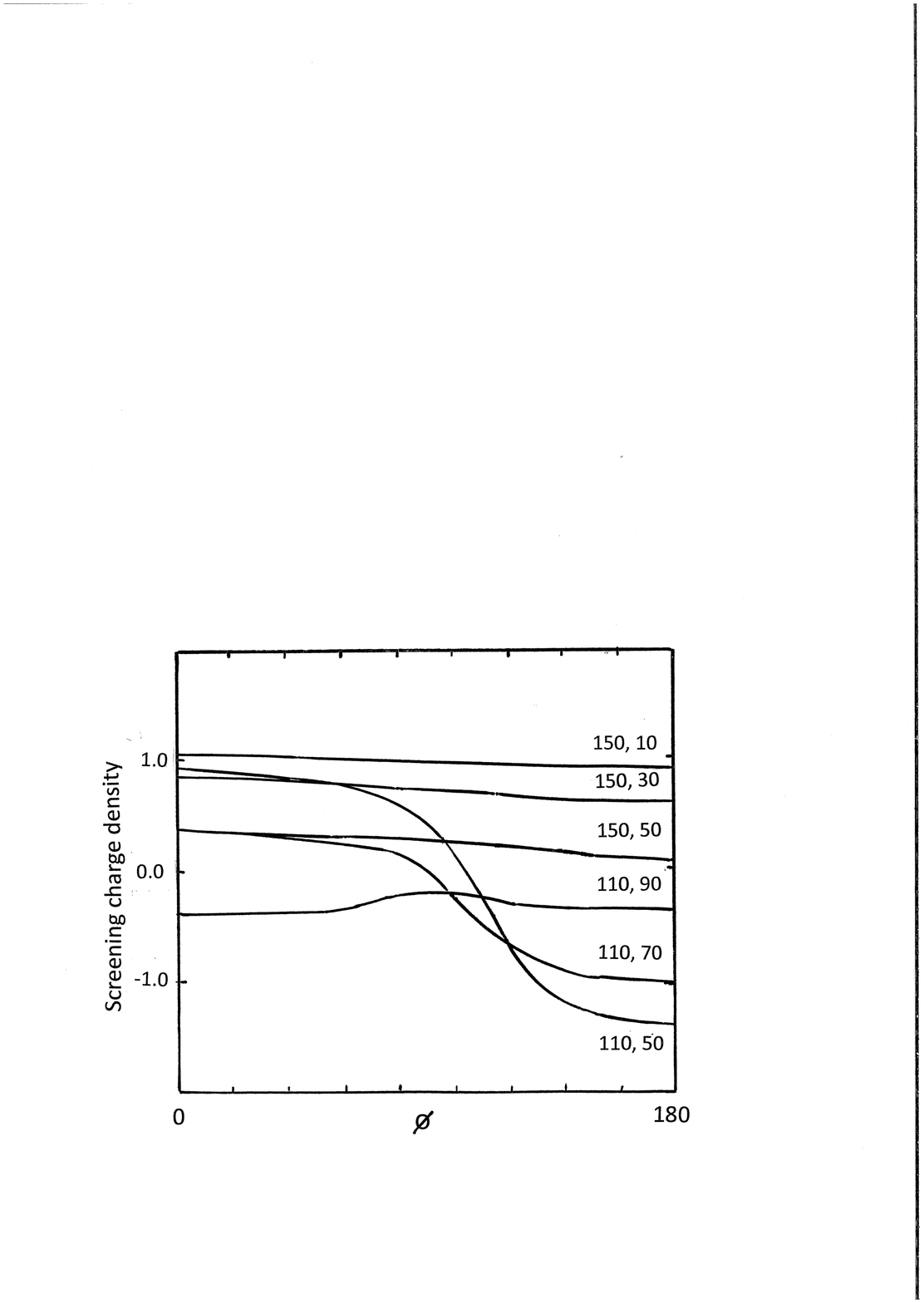}

\caption{The screening charge density $\rho_{s}$ is stationary in the corotating frame but must be supplied by the drifting plasma.  It is shown here in units of $\rho_{GJ0}$ at $t = 0$ as a function of the azimuthal angle $\phi$ for fixed values of the obliquity $\psi$ and polar angle $\theta$.  The current flow from the neutron-star surface required to maintain the correct charge density is determined by the rate of variation of $\rho_{s}$ as a function of $\phi$.  Each relation is labelled $(\psi, \theta)$ in degrees.}
\end{figure}

In either case a reverse transfer of protons must also be present, but here the very great mass difference becomes significant in that the electrons move with relativistic speeds at quite low levels of $E_{\parallel}$.  Compensation then occurs on any flux line at a rate limited only by the velocity of light, as described by Melrose \& Yuen (2014).  In the absence of an induction field, there would be a problem in understanding how the outward flux of electrons could move perpendicular to ${\bf B}$ and so be enabled to cross the light cylinder.  The presence of the drift velocity $\tilde{\bf v}$ solves this directly.  In the following Section we attempt to describe the effect of drift into the open magnetosphere.

\section{The polar cap and open sector of the magnetosphere}

We shall assume that the open magnetosphere divides into two sectors as first envisaged by Arons \& Scharlemann (1979).  Its precise definition requires a demarcation between those particles that cross the light cylinder and those that do not: it depends on knowledge of the ${\bf E}$ and ${\bf B}$ fields near the light cylinder.  Numerical plasma physics, such as the work of Bai \& Spitkovsky (2010) in the force-free approximation, indicates that the cross-sectional area of the open magnetosphere is roughly circular at the inner radius of $0.2\eta_{LC}$ adopted by them. We refer to flux lines on which the effect of the polar-cap acceleration field first described by Muslimov \& Tsygan (1992) is not cancelled by flux-line curvature at higher altitudes as the active sector within which the energy of outward-accelerated particles is the source of the coherent radio emission. For favourable flux-line curvature, with $\sigma = \eta^{3}\rho_{s}(\eta)$ an increasing function of $\eta$, equation (6) shows that further acceleration occurs beyond that of the Lense-Thirring effect. The remaining open flux lines will be referred to as the return sector.  This is obviously not a precise definition, but those lines on which a zero in ${\bf \Omega}\cdot{\bf B}$ exists inside the light cylinder are certainly in the return sector.  This division is shown in Fig. 2 which also shows the drift relative to the boundary of the open field lines as it would be if there were no inductive field. To be specific, we adopt
a circular form for the open magnetosphere boundary ${\bf u}_{0}$ with radius $u_{0} = 1.6\times 10^{4}P^{-1/2}$ cm for a $1.4$ $M_{\odot}$ star (Harding \& Muslimov 2001).  Above the polar cap and for dipole geometry, the open magnetosphere consists of a long narrow tube of flux lines whose cross-sectional area given by the open sector radius $u_{0}(\eta) = \eta^{3/2}u_{0}(1)$ varies little over lengths of the order of the neutron-star radius.  Thus if there were no division into active and return sectors, the potential at any point inside the tube away from the polar-cap surface could be expressed as,
\begin{eqnarray}
\Phi({\bf u},\eta) = \pi \left(u_{0}^{2} - u^{2}\right)\left(\rho - \rho_{s}\right).
\end{eqnarray}
This elementary approximation is adequate, particularly as the reduction in cross-sectional area caused by the division into active and return sectors is uncertain and reduces the potential, $\Phi \rightarrow \zeta\Phi$, where $1/4 < \zeta <1/2$ for a semi-circular shape. Equation (6) assumes the condition $\Phi = 0$ on the open-closed boundary and that it is maintained in the presence of drift.

\begin{figure}
\includegraphics[trim =10mm 40mm 20mm 100mm, clip, width=84mm]{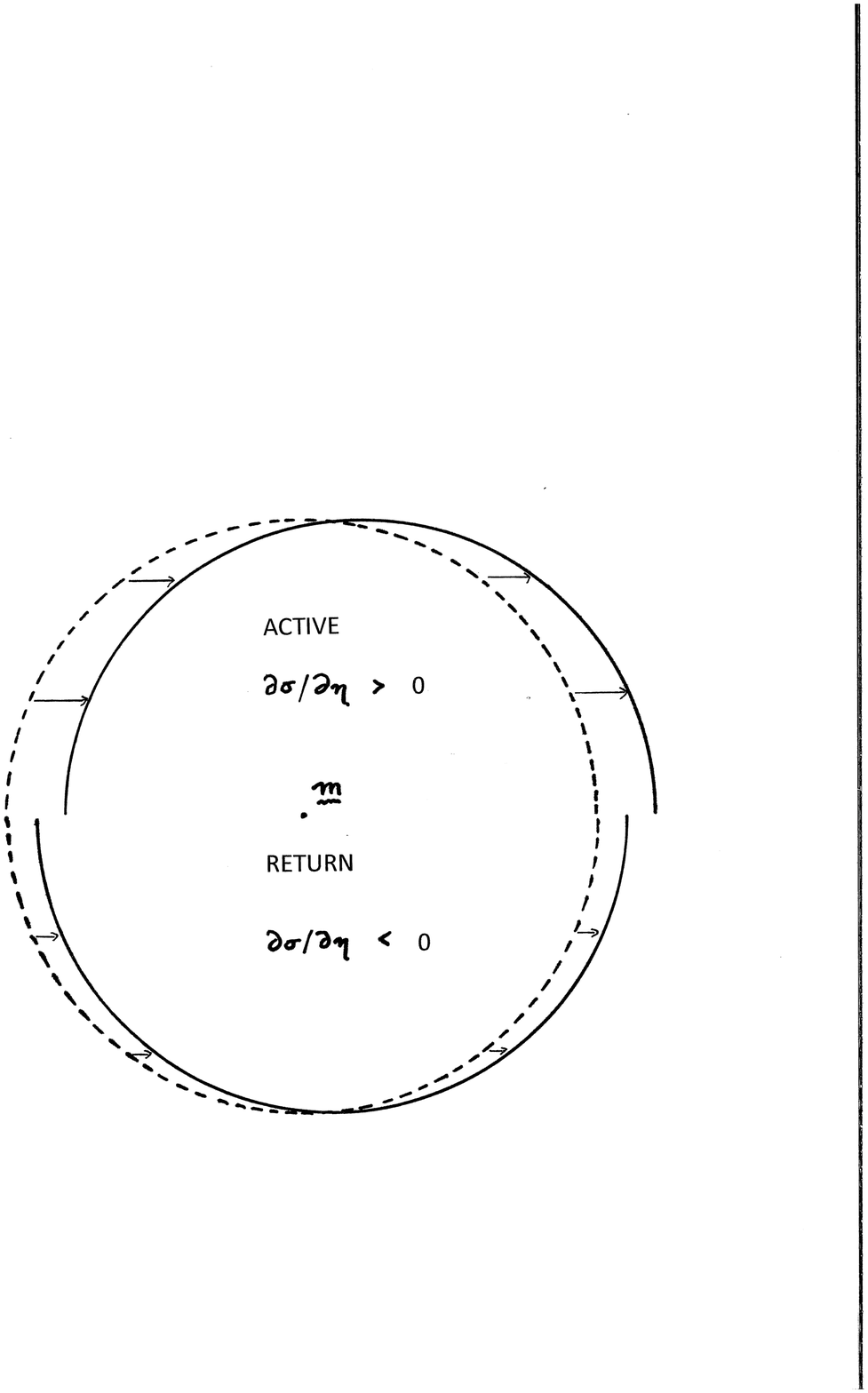}

\caption{The cross-section of the open magnetosphere at an altitude $\eta = 100$ is shown (broken line) divided (equally) between the active and return sectors and centred on the magnetic axis ${\bf m}$  as it would be in the absence of drift.  Its radius (for dipole geometry) is $1.6\times 10^{7}P^{-1/2}$ cm.  The displacement of particles accelerated from the neutron star surface in the active sector is shown by the displaced semi-circle and is $3\times 10^{6} P^{-1}$ cm on the basis of equation (7).  The displacement of particles in the return sector which are not accelerated parallel with ${\bf B}$ is also shown and is one half of that in the return sector.  The discontinuity shown is a result of our approximations and has no consequences of significance.}
\end{figure}

The induction field component ${\bf E}^{ind}_{\perp}$ causing drift is present in the open sector, and its effect on particle motion follows at once by considering motion parallel with, and perpendicular to, the flux lines.  On the basis of equation (4), a particle entering an active dipole-field sector at altitude $\eta_{1}$ drifts a distance,
\begin{eqnarray}
b_{\perp} & = & \frac{R}{v_{\parallel}} \int^{\eta_{2}}_{\eta_{1}} d\eta \tilde{v}_{\phi}\left(\frac{\eta_{2}}{\eta}\right)^{3/2}   \nonumber \\
  & = & H\eta_{2}^{3/2}\left|\eta_{2}^{1/2} - \eta_{1}^{1/2}\right|,
\hspace{1cm} H = \frac{\Omega R^{2}\sin\psi}{v_{\parallel}}
\end{eqnarray}
perpendicular to the flux line it intersected at entry whilst moving, inward or outward, with velocity $v_{\parallel}$  to $\eta_{2}$.  The basic unit of length is $H = 300P^{-1}\sin\psi$ cm  for $R = 1.2\times 10^{6}$ cm and $v_{\parallel} = c$, where $P$ is the rotation period in seconds. The velocity in equation (7) is the instantaneous $\tilde{v}_{\phi}$, not its average over a rotation period.

With this evaluation, it is easy to see that drift of protons or ions into the active open sector from the closed has little effect on the polar-cap physics of normal pulsars at altitudes less than $\eta_{1} \approx 10$ which contain the source of coherent radio emission (Hassall et al 2012).  In this case, the drift $b_{\perp}$ at $\eta_{2} \approx 10$ is small compared with the width of the active flux tube and the generation of radio frequencies is not impaired.  This is illustrated by the example shown in Fig. 2 {\bf for $\sin\psi = 1$.}   However, for $\eta_{2} \gg 10$, it is possible that a fraction of protons accelerated from the neutron-star surface will enter the closed sector, but the immediately observable consequences of this, if any, are unclear.  However, in the case of millisecond pulsars (MSP), the drift is much more significant, although the light-cylinder radius is small, $\eta_{LC} \sim 20$. Even so, ion and proton acceleration remains consistent with radio emission spectra broadly of the universal form, as is observed (Jenet et al 1998, Kramer et al 1999, Espinoza et al 2013). The direction of drift is such that $\tilde{v}_{\phi} > 0$, that is, super-rotation.  It is possible to confirm, directly from equation (10) of Melrose \& Yuen (2012), that its direction is identical for both signs of ${\bf \Omega}\cdot{\bf B}$.

Drift into the return open sector is a different matter. In the absence of an induction field, initial acceleration of a test particle arising from the Lense-Thirring effect would be nullified in equation (6) by the unfavourable flux-line curvature, for which $\sigma(\eta)$ is a decreasing function of $\eta$. Given free transit of protons (or ions) at the neutron-star surface, the consequence would be stasis unless secondary electron-positron plasma exists whose overall charge density could adjust so as to accommodate to a null point and an adverse flux-line curvature.  But in the absence of pairs, input of protons from the positive sector of the closed magnetosphere merely contributes to this state and drift continues azimuthally through the return open sector into the closed.

Input of electrons from the closed sector beyond the null point on any flux line is followed by outward acceleration to the light cylinder.  Melrose \& Yuen caution that their published $\tilde{\rho}$ and $\tilde{\bf v}$ may be unreliable near the light cylinder, but we shall use them to estimate that the total electron current passing from the closed sector is of the order of,
\begin{eqnarray}
\int^{\eta_{LC}}_{\eta_{null}} d\eta Ru_{0}(\eta)\rho_{GJ0}(\eta)\tilde{v}_{\phi}   \nonumber  \\
 = Hcu_{0}(1)\rho_{GJ0}(1)\left(\eta^{1/2}_{LC} - \eta^{1/2}_{null} \right),
\end{eqnarray}
estimated using the instantaneous $\tilde{v}$.   We can see that $H \eta_{LC}^{1/2} = R\sin\psi \eta^{-1/2}_{LC}$ is of the same order as $u_{0}(1)$ so that, qualitatively, the electron current given by equation (8) is adequate to compensate for the outward flux of  positive charge.

\section{The outer gap}

Although the non-corotation may not be exactly as described by Melrose \& Yuen, it does have implications for the outer-gap model introduced by Cheng, Ho \& Ruderman (1986) who first noticed the possible significance of the null surface in what we have termed the return sector of the magnetospheric charge density distribution. In the ${\bf \Omega}\cdot{\bf B} < 0$ case, electrons injected beyond it are accelerated outward to high energies at the light cylinder and could, in principle, be the source of such incoherent radiation as is observed.   But finding the source of the electrons, or of electron-positron pairs, has always been a problem. Single-photon pair production by curvature radiation is not usually possible and so a further source, photon-photon collisions of curvature photons and polar-cap blackbody radiation was later introduced by Cheng, Ruderman \& Zhang (2000).

There has been much work on outer-gap models of $\gamma$-emission in the MSP and other pulsars, but particularly in the MSP, the above problem remains. The non-corotation flux of electrons we have referred to in the final paragraph of the previous Section is, in principle, a solution.  Electron drift into this sector occurs naturally in the ${\bf \Omega}\cdot{\bf B} < 0$ case and the  integrated flux estimated by equation (8) is of the same order of magnitude as the Goldreich-Julian active-sector flux from the polar cap. It is known that high-energy electrons (or positrons) must be present in the vicinity of the light cylinder because MSP feature numerously in the Fermi LAT catalogue of $\gamma$-emitting pulsars (Abdo et al 2013). But it is extremely unlikely that the formation of a dense plasma of outward-moving secondary pairs by curvature radiation is possible in the outer gap in any model.  It must be presumed that the outward-moving particles are exclusively electrons and that Goldreich-Julian fluxes are sufficient to produce the observed $\gamma$-ray spectrum. A further feature of the catalogue is the absence of any radio-quiet MSP (defined as a flux density $S_{1400} < 30$ $\mu$Jy) which is considered notable even though their detection procedures are necessarily different.  If the outer gap is the source of $\gamma$-emission in the MSP, the neutron star must have polar-cap ${\bf \Omega}\cdot{\bf B} < 0$. This is consistent with the source of the universal radio emission being an ion-proton plasma at an ${\bf \Omega}\cdot{\bf B} < 0$ polar cap (see Jones 2015). We do not expect radio-frequency emission from an ${\bf \Omega}\cdot{\bf B} > 0$ polar cap and in such a neutron star, particle drifting into the outer gap region would necessarily be positively charged, that is, ions or protons, whose acceleration would produce negligible incoherent emission.

\section{Conclusions}

Although the arguments of Melrose \& Yuen (2014) are persuasive, these authors bear in mind the relative ubiquity of corotation and suggest that it might be enforced by some mechanism as yet unknown. But we have discounted this possibility and in this paper have taken their results for screening charge density and drift velocity as correct, although this cannot be so near the neutron-star surface because, as they point out, they assume a point-dipole field or near the light cylinder (for example, the denominator term in the Goldreich-Julian charge density has been neglected).  The case in which ${\bf \Omega}\cdot{\bf B} < 0$ has been chosen for discussion because we consider that such neutron stars are the sources of the almost universal coherent emission that is observed whether in normal pulsars or in MSP (Jones 2013, 2015).  The fact that screening of the induction field component ${\bf E}^{ind}_{\parallel}$ may be incomplete at radii near the light cylinder is likely to have interesting consequences for particle acceleration in that region, as noted Melrose \& Yuen (2012).

In relation to the latter comment, we note that electron drift into the region of the return sector beyond the null surface at once solves the problem of how the outer gap is populated. Observable incoherent emission, particularly $\gamma$-emission is then possible in neutron stars which are not capable of supporting pair creation.

But the polar cap in a magnetosphere that is not in strict corotation has, in principle, properties which are unchanged insofar as radio emission is concerned, certainly if the source is at altitudes $\eta \sim 10$ as favoured by Hassall et al (2012).  The presence of drift relative to the corotating frame also solves, in principle, the return current problem.

It follows that there appear to be no obvious observational consequences of non-corotation for radio emission.  The closed sector is in a dynamic state.
Time-dependent currents needed to maintain the screening $\rho_{s}$ and the open-closed boundary condition are always present. Particle flow inward to the atmosphere is likely to be dissipative but the local heating arising from a Goldreich-Julian flux distributed over much of the neutron-star surface is unlikely to be observable.

The possibility that counter-flowing beams of oppositely charged particles might co-exist in a more detailed description of non-corotation is of interest in the case of ${\bf \Omega}\cdot{\bf B} > 0$ neutron stars.  For an ion-proton beam with counter-flowing electrons, Langmuir modes exist whose wave-vectors are parallel with the ion-proton velocity and so, for this spin sense, probably directed outward towards the observer.  They require both beams to have quite small Lorentz factors, but the kinematics of the process allows the generation of substantial radio-frequency power. We refer to Jones (2014) for the methods of analysis which lead to this conclusion. The observation of unusual or low flux-density sources in the Square Kilometre Array (SKA) would be interesting in this context.

\section*{Acknowledgments}

I thank the anonymous referee for some very helpful recommendations which have improved the presentation of this work.

\bsp

\label{lastpage}

\end{document}